\documentstyle[aps,epsf,rotate,preprint]{revtex}

\newcommand{\sphalf}{1.35}

\newcommand{\kimspace}{\edef\baselinestretch{\sphalf}\Large\normalsize}

\def\be{\begin{equation}}
\def\ee{\end{equation}}
\def\ba{$\begin{array}{c}}
\def\ea{\end{array}$}

\def\a{A}
\def\b{B}
\def\c{C}
\def\o#1{{\cal O}(\lambda^#1)}
\def\l#1{\lambda^#1}
\def\s2{\sqrt{2}}

\begin{document}


\author{T.~K.~Kuo\thanks{tkkuo@physics.purdue.edu} , 
Sadek~W.~Mansour\thanks{mansour@physics.purdue.edu}, 
and Guo-Hong Wu\thanks{wu@physics.purdue.edu}\\
{\it Department of Physics, Purdue University,\\ 
\it West Lafayette, Indiana 47907}
}
\title{Classification and Application of Triangular Quark Mass Matrices}
\date{Revised Version, September 1999}
\maketitle   
\begin{abstract}

 Any given pair of up and down quark mass matrices can be
brought into the triangular form, at the same time
eliminating their unphysical contents.  Further, we show that every
physically viable pair can always be reduced to one of ten 
triangular forms, which we list explicitly.
When hermitian mass matrices are thus analyzed, it is found that
any pattern of texture zeros translates into simple, analytic
predictions for the quark mixings.  
Amongst possible hermitian matrices with five texture zeros, 
this method enables us to identify
the unique pair which is favored by present data.

\end{abstract}


\newpage

{\bf 1.~}{\bf Introduction.} 

In the study of quark mass matrices, one often starts from  
a presumed form and then derives relations 
between the masses and mixings that can be confronted with 
experimental data. Usually, the forms considered are hermitian 
with certain texture zeros~\cite{Fritzsch,RRR}, although non-hermitian matrices have
also been considered~\cite{Branco}. 
It has recently been observed \cite{KMW} that upper triangular 
matrices provide a most convenient basis to get the physical
contents of general quark mass matrices.  
Through a rotation of the right-handed (RH) quarks,
any mass matrix (defined as in $\overline{\psi_L} M \psi_R$) 
can be put into the upper triangular form, with 
three zero elements at the lower-left corner.
When either the up or the down quark mass matrix is diagonal, 
the triangular mass matrix elements of the other
consist simply of products of a mass and a Cabbibo-Kobayashi-Maskawa (CKM)
matrix element.
In particular, the diagonal elements are approximately equal to the 
quark masses, and the off-diagonal elements directly tell us 
about the mixing among the three families.
$CP$-violating effects depend on one independent phase that can 
reside in any of the $(1,2)$, $(1,3)$, $(2,2)$, and $(2,3)$ 
positions of the triangular matrix.
It is obvious that the total number of parameters in the mass matrices
is equal to the number of physical observables, namely, 
the six quark masses, three mixing angles, and one $CP$ phase.
In fact, among the existing minimal parameter bases \cite{mpb,Haussling}, 
where the unphysical degrees of freedom are eliminated,
upper triangular mass matrices exhibit most clearly and simply 
the quark masses and CKM angles.

The purpose of this letter is two-fold. 
Firstly, we would like to extend our previous analysis to the more 
general case where neither the $U$- nor the $D$-quark mass matrix is diagonal.
In particular, we classify all the upper triangular mass matrices
in the minimal parameter basis -- with nine nonzero elements distributed
between the up and down quark mass matrices. 
The observed hierarchical structure of quark masses and small mixing angles
allows us to choose the hierarchical basis for the mass matrices in which 
the $(3,3)$ is the largest element and the rotation angles needed 
for diagonalization are all small. 
In this basis, we obtain a complete set of ten pairs of upper triangular 
quark mass matrices. To a good approximation, each entry 
is simply given by either a quark mass or its product with one or two 
CKM elements. 
Any given pair of mass matrices can always be reduced to one of the
pairs listed in Table I. In so doing, one can easily read off
the physical masses and mixings implied by these matrices.

  Secondly, we use triangular matrices to study texture zeros in
hermitian mass matrices which may be a manifestation of certain flavor
symmetries.
It is found that simple relations can be established between the 
parameters of hierarchical hermitian and triangular matrices.
By reducing generic hermitian matrices into one of the pairs
in Table I, we show that any given pattern of texture zeros
in the hermitian matrices implies simple relations between quark
masses and mixing angles.
Specifically, analyses of these relations show that 
only one of the five pairs of hermitian matrices
studied by Ramond-Roberts-Ross (RRR) in Ref.~\cite{RRR}
is favored by current data.

{\bf 2.~}{\bf Classification.} 

We start by deriving the triangular form of the down (up) quark mass matrix
in the basis where $M^U$ ($M^D$) is diagonal. 
Then we generate all the other triangular
textures in the minimal parameter basis through a common left-handed (LH)
rotation on both $M^U$ and $M^D$ together with separate RH rotations, 
when necessary.  
Without loss of generality,  we will work in a basis where 
${\rm det}V_{\rm CKM}=+1$,
which will simplify the expressions for the mass matrices.

For $M^U={\rm Diag}(m_u,m_c,m_t)$, $M^D$ is given by
$M^D=V_{\rm CKM}{\rm Diag}(m_d, m_s, m_b)$ up to a RH unitary
rotation. By applying to $M^D$ three successive RH rotations,
we can bring the $(3,1)$, $(3,2)$, and $(2,1)$ elements of $M^D$
to zero and arrive at the upper triangular form\cite{KMW},
\be \label{eq1} 
M^D  = \left(\begin{array}{ccc}
\frac{m_d}{V_{ud}^*}&m_sV_{us}&m_bV_{ub}\cr 
0&m_sV_{cs}&m_bV_{cb}\cr 
0&0&m_bV_{tb} \end{array}\right) \times (1+\o4) \; ,\;\;\;\;\;\;\;\; (M^U~{\rm diagonal})
\ee
where $\lambda=|V_{us}|=0.22$~\cite{Wolfenstein}.

Eq.~(\ref{eq1}) is almost the same as $V_{\rm CKM}{\rm Diag}(m_d, m_s, m_b)$ by
striking out the three lower-left matrix elements. The reason for this
simple form is because the afore-mentioned RH rotations are all
very small, being of order $\lambda^7$, $\lambda^4$, and $\lambda^3$ for the 
$(31)$, $(32)$, and $(21)$ rotations, respectively. The most noticeable effect of 
these comes from the $(21)$ rotation which changes the $(1,1)$ element 
of $M^D$ by $1+\o2$.
The factor $1/V_{ud}^*$ arises because of the invariance of the
determinant, and because $V_{ud}^*=V_{cs}V_{tb}-V_{cb}V_{ts}\simeq V_{cs}V_{tb}$,
 for ${\rm det}V_{\rm CKM}=+1$. 
Note that, by transforming a hierarchical matrix 
into the upper triangular form, one has eliminated the RH rotations necessary for its diagonalization.

By contrast, lower triangular mass matrices~\cite{Haussling}, with
zeros in the upper-right corner, do not provide a simple relation between its 
matrix elements and the quark masses and CKM angles. The reason is that, had we
tried to make the matrix $M^D=V_{\rm CKM}{\rm Diag}(m_d, m_s, m_b)$ lower 
triangular, we would have to use large angle RH rotations. In doing so,
the resulting lower triangular matrix elements are very different from those of
the original matrix, and no simple relations are expected.

When $M^D$ is diagonal, the upper triangular form
for $M^U$ can be simply obtained from Eq.~(\ref{eq1}) by 
replacing $V_{ij}$ with $V_{ji}^*$~\cite{KMW},
\begin{eqnarray} \label{eqMU}
M^U  & = & \left(\begin{array}{ccc}
 \frac{m_u}{V_{ud}} &m_cV_{cd}^*&m_tV_{td}^*\cr
0&m_cV_{cs}^*&m_tV_{ts}^*\cr 
0&0&m_tV_{tb}^* \end{array} \right) \times (1+\o4) \; .\;\;\;\;\;\;\;\; (M^D~{\rm diagonal})
\end{eqnarray}
Eqs.~(\ref{eq1}) and (\ref{eqMU})
 will serve as our starting point for getting all the other 
triangular mass matrices with nine nonzero elements.
To retain a simple expression for them in terms of
physical parameters, we will work in the hierarchical basis 
in which the magnitudes of the matrix elements, if non-zero, satisfy
$M_{33} \gg M_{22} \gg M_{11}$, $M_{33} \gg M_{23} \gg M_{13}$
 and  $M_{22} \gg M_{12}$. 

  To generate other triangular textures from Eq.~(\ref{eq1}), 
we can set one of the off-diagonal elements in $M^D$ to zero through
a LH $D$-quark rotation (more precisely, this is the leading order
term of a succession of LH and RH rotations with decreasing
angles). The same LH rotation acting on $M^U$ 
then brings the diagonal $M^U$ into upper triangular form with
one non-zero off-diagonal element.  
To remain in the hierarchical basis, only
small angle (smaller than $\pi/4$) LH rotations 
are allowed.  
We thus obtain the first five pairs of matrices as listed in Table~\ref{table1}.
From Eq.~(\ref{eqMU}), another set of five pairs of triangular matrices can be
obtained.
They are related to the first five in Table~\ref{table1} 
by the operation:
$M^U \leftrightarrow M^D$ and $V_{ij} \leftrightarrow V^*_{ji}$.
Note that the texture zeros are phenomenological zeros, with 
negligible but nonvanishing physical effects.
In this way, we arrive at a complete list of ten pairs of upper triangular
textures in the hierarchical, minimal parameter basis. 

For the textures of Table~\ref{table1}, the CKM matrix can be written, 
to a good approximation, as a product of 
three LH rotations, coming separately from the diagonalization of $M^U$ and $M^D$.
The rotation angles are approximately given by the ratios of CKM elements.
These are also listed in Table~\ref{table1}.
Finally, for each pair of textures in the minimal parameter basis, 
there is only one physical phase which can be written as  a linear combination 
of either four or six phases of the matrix elements. 
This phase is simply related to one of the three
angles of the unitarity triangle:
$\alpha \equiv {\rm arg}\left[-\frac{V_{td} V_{tb}^*}{V_{ud} V_{ub}^*}\right]$,
$\beta \equiv {\rm arg}\left[-\frac{V_{cd}V_{cb}^*}{V_{td} V_{tb}^*}\right]$,
and
$\gamma \equiv {\rm arg}\left[-\frac{V_{ud} V_{ub}^*}{V_{cd} V_{cb}^*}\right]$.
It enters into the $CP$-violating
Jarlskog parameter~\cite{Jarlskog}, the approximate form of which is also
listed in Table~\ref{table1}.

The simplicity of the triangular mass matrices in Table~\ref{table1} is
evident: each matrix element is either a quark mass or its product with
one or two CKM elements.
By contrast, other minimal-parameter textures cannot
be written in such a simple form.
Table~\ref{table1} allows one to read off immediately the physical 
masses and mixings of any quark mass matrices after casting them 
into one of the ten triangular patterns. 
The conversion can be achieved through separate RH rotations and possibly 
LH rotations which are common to both $M^U$ and $M^D$.
This could avoid the sometimes tedious process of matrix diagonalization~\cite{KMW}.

{\bf 3.~}{\bf Triangular versus hermitian mass matrices.}

  One can equally well start from upper triangular mass matrices, 
in which the unphysical RH rotations are eliminated,
and generate other forms of mass matrices by appropriate RH rotations. 
In particular, we may generate hermitian mass matrices, which have been 
the focus of many studies. 
Here we illustrate the method by examining analytically hermitian 
mass matrices with five texture zeros, as was studied in \cite{RRR}.

   Using the mass relations $m_u:m_c:m_t \sim \lambda^8:\lambda^4:1$
and $m_d:m_s:m_b \sim \lambda^4:\lambda^2:1$ evaluated at the weak scale, 
and the CKM rotations
$V_{us} = \lambda$, $V_{cb} \sim \lambda^2$, 
$V_{ub} \sim \lambda^4$, and $V_{td} \sim \lambda^3$, we can write the 
properly normalized Yukawa matrices for $U$ and $D$
in the most general triangular form,
\begin{equation} \label{eq:triUD}
T^U = \left(\begin{array}{ccc}
       a_U \lambda^8 & b_U \lambda^6 & c_U \lambda^4 \\
          0          & d_U \lambda^4 & e_U \lambda^2 \\
          0          &    0          &  1            
           \end{array}\right)\; ,
\;\;\;\;\;\;\;\;\;\;
T^D = \left(\begin{array}{ccc}
       a_D \lambda^4 & b_D \lambda^3 & c_D \lambda^3 \\
          0          & d_D \lambda^2 & e_D \lambda^2 \\
          0          &    0          &  1            
           \end{array}\right)\; ,
\end{equation}
where the diagonal coefficients are determined by quark masses and are
 of order one, and the off-diagonal coefficients can be either of order one 
or much smaller.
Without loss of generality, we can take the diagonal elements to be real
in the following analysis.
In writing the above triangular matrices, we have required 
the LH $(12)$, $(23)$, and $(13)$ rotations 
for diagonalizing each sector to be respectively of order $V_{us}$, $V_{cb}$, 
and $V_{td}$ ($V_{ub}$) or smaller\footnote{Though we take the LH $(12)$ 
rotation in the $U$ sector to be of order  $\sqrt{m_u/m_c} \sim \l2$
for convenience, one can also start with  
$T^U_{12}=b_U \lambda^5$ and arrive at the same results.}.
This is known as the naturalness criteria~\cite{PW}, and can always be 
implemented by a common LH rotation to both $T^U$ and $T^D$.

  From the hierarchical form of Eq.~(\ref{eq:triUD}), one can directly 
read off both the mass eigenvalues and the LH unitary matrices 
that diagonalize $T^U$ and $T^D$.
The quark masses are simply related to the diagonal elements, 
\begin{eqnarray} 
 (m_u, m_c) & = & m_t(a_U \lambda^8, d_U \lambda^4) \times (1+\o4)\; ,  \\
 (m_d, m_s) & = & m_b(a_D \lambda^4, d_D \lambda^2) \times (1+\o2)\; .
\end{eqnarray}
The CKM matrix is obtained from separate LH diagonalizing rotations for $T^U$ and $T^D$~\cite{KMW},
\begin{eqnarray}
V_{\rm CKM} & = & V_U^{\dag} V_D \; , \nonumber \\
V_U & \simeq & R_{23}(e_U \lambda^2) R_{13}(c_U \lambda^4) 
               R_{12}(b_U \lambda^2/d_U ) \; , \nonumber \\
V_D & \simeq & R_{23}(e_D \lambda^2) R_{13}(c_D \lambda^3)
               R_{12}(b_D\lambda/d_D ) \; . 
\end{eqnarray}

   The mass matrices can take different but physically
equivalent forms through RH rotations.
In particular, they can be transformed into the following hermitian form
($T^i\rightarrow Y^i\simeq T^i V_i^{\dag}$, $i=U,D$),
\begin{eqnarray} \label{eq:HU}
Y^U  & = & \left(\begin{array}{ccc}
       \left(a_U +  c_U c_U^* + 
               \frac{b_Ub_U^*}{d_U}  \right) \lambda^8 & 
       (b_U + c_U e_U^*)  \lambda^6 & c_U \lambda^4 \\
       (b_U^* + c_U^* e_U)  \lambda^6  & 
       (d_U + e_U e_U^*) \lambda^4 & e_U \lambda^2 \\
          c_U^* \lambda^4  & e_U^*\lambda^2   &  1            
           \end{array}\right) \times (1+\o4)\; ,
\end{eqnarray}
\begin{eqnarray} \label{eq:HD}
Y^D & = & \left(\begin{array}{ccc}
      \left(a_D +  \frac{b_Db_D^*}{d_D} \right) \lambda^4 & 
       b_D \lambda^3 & c_D \lambda^3 \\
       b_D^* \lambda^3  & d_D \lambda^2 & e_D \lambda^2 \\
        c_D^* \lambda^3  &    e_D^* \lambda^2     &  1            
           \end{array}\right) \times (1+\o2) \ .
\end{eqnarray}

   A few interesting observations can now be made about $Y^U$ and $Y^D$. 
\begin{enumerate}
\item {\em hermitian vs. triangular zeros:}
Except for $Y^U_{12}$, it is seen that an off-diagonal zero\footnote
{Only zeros along and above the diagonals of triangular 
and hermitian matrices are counted.} in the triangular form has a one to one correspondence to that in the hermitian form. On the other hand, whereas diagonal zeros are not allowed in the 
triangular form (corresponding to vanishing quark masses), for $Y^U$ and $Y^D$
such zeros imply definite relations among elements of the triangular matrix. 
\item {\em $Y^U$ vs. $Y^D$ zeros:}
A notable difference between $Y^U$ and $Y^D$ is that 
$Y^U_{22}$ can be zero, but $Y^D_{22} \neq 0$\footnote{More precisely,
$Y_{22}^D=(d_D+|e_D|^2\lambda^2)\lambda^2$ so that 
$Y^D_{22}=0$ would imply $\theta^D_{23}\simeq \sqrt{m_s/m_b}\sim \lambda$,
which is too large to be partially canceled by $\sqrt{m_c/m_t}$ to 
get $V_{cb} \sim \lambda^2$. Had the cancellation somehow occurred,
it would involve ${\cal O}(\lambda)$ fine tuning which is against 
the naturalness criteria. }.
$Y^D_{11}=0$ implies that the (12) rotation angle for $Y^D$ is $\theta_{12}^D\simeq 
\sqrt{m_d/m_s}$. 
The condition $Y^U_{22}=0$ implies $\theta^U_{23} \simeq \sqrt{m_c/m_t}$. 
On the other hand, $\theta^U_{12} \simeq \sqrt{m_u/m_c}$ can be
obtained if both $Y^U_{11}=0$ and $Y^U_{13}=0$ (more precisely
$|c_U^2/a_U| \ll 1$ ) are satisfied.
 One other relation, $\theta^U_{23} \theta^U_{13}/\theta^U_{12}
 \simeq m_c/m_t$, can be obtained if $Y^U_{12}=0$.
\item
$Y^D_{11}$ and $Y^D_{12}$ cannot be both equal to zero.
One can therefore have at most 3 texture zeros 
for $Y^D$ corresponding to 
$Y^D_{11}=Y^D_{13}=Y^D_{23}=0$ or $Y^D={\rm Diag}(m_d/m_b, m_s/m_b,1)$.
Similarly, $Y^U_{22}$ and $Y^U_{23}$ cannot be both zero,
and one can easily show that at most 3 texture zeros are allowed
for $Y^U$ with five possible forms including the diagonal matrix.
\item
The zeros in $Y^U$ and $Y^D$, which imply relations between 
the elements of the triangular matrices,
translate into relations between physical parameters by referring to 
Table~\ref{table1}.
We can thus quickly rule out pairs of $Y^U$ and $Y^D$'s which are obviously
untenable. It is easy to check that there are no viable pairs with six 
texture-zeros, and that only five pairs with five texture zeros are possible
candidates, in agreement with the findings of Ref.~\cite{RRR}. However, the triangular matrix 
method also allows us to investigate each pair analytically, and to have 
a more in-depth evaluation of the physical implications of these matrices. 
\end{enumerate}

In Table~\ref{table2}, we list five types of hermitian matrices, 
together with their corresponding triangular forms. 
These five hermitian matrices can be paired up to give
the five RRR patterns for the quark Yukawa matrices, $(Y^U,Y^D)$,
which are listed in Table~\ref{table3}. 
Using Tables \ref{table2} and \ref{table3}, we see that RRR patterns 1, 3, 
and 4 simply correspond to triangular textures II, III, and VII of 
Table~\ref{table1}, respectively. 
The triangular form of RRR pattern 2, $(M_2, M_4)$, has only two 
texture zeros, and an appropriate LH (23) rotation can be used to generate a third zero
and transform it into either triangular patterns II or VII. 
Similarly, a LH (12) rotation is necessary to bring RRR pattern 5,
$(M_5, M_1)$, into the minimal-parameter triangular patterns VI, VII, or X. 
The physical implications are summarized in the last column of 
Table~\ref{table3}. For instance, the relation $|b_U|^2=-a_Ud_U$ for pattern 1,
referring to texture II of Table~\ref{table1}, reads 
$\left|m_c \frac{V_{ub}}{V_{cb}}\right|^2 =|m_u m_c|$, 
or $\left|\frac{V_{ub}}{V_{cb}}\right|^2 = |\frac{m_u}{m_c}|$.
Each of the five RRR patterns entails two physical predictions\footnote
{For patterns 2 and 5, because of the necessary LH rotation, 
the three relations listed also give rise to only two physical predictions.}.
We turn now to a detailed analysis of these predictions.

   From Tables~\ref{table1} and~\ref{table3}, RRR patterns 1, 2, and 4 
give rise to the same predictions:
\begin{eqnarray}
\left| \frac{V_{ub}}{V_{cb}} \right| & = & \sqrt{\frac{m_u}{m_c}} \times (1+\o3)
 = 0.059 \pm 0.006 \; , 
    \label{eq:Vubcb} \\
\left| \frac{V_{td}}{V_{ts}} \right| & = & \sqrt{\frac{m_d}{m_s}}\times (1+\o2)
 = 0.224 \pm 0.022   \label{eq:Vtdts} \; ,
\end{eqnarray}
where the quark masses are taken from Ref.~\cite{RGE}.
It is interesting to note that the ratios of the quark masses and mixings
in Eqs.~(\ref{eq:Vubcb}) and (\ref{eq:Vtdts}) are all 
scale-independent~\cite{OlePor,RRR}. 
Furthermore, Eqs.~(\ref{eq:Vubcb},\ref{eq:Vtdts}) are independent of the 
phases in the mass matrices.
Experimentally,
\begin{eqnarray} \label{eq:exVubVtd}
\left| \frac{V_{ub}}{V_{cb}} \right|_{\rm exp} &=& 0.093 \pm 0.014 \; ,\\
0.15 &<& \left| \frac{V_{td}}{V_{ts}} \right|_{\rm exp} < 0.24 \; ,
\end{eqnarray}
where $V_{ub}$ is from a recent measurement~\cite{Parodi,Ali} and $\left|\frac{V_{td}}{V_{ts}}\right|$ comes 
from a 95\% C.L. standard model fit to electroweak data~\cite{Ali}.
We may conclude that the 1st, 2nd, and 4th RRR patterns
are disfavored by the data on $V_{ub}/V_{cb}$.

   Tables~\ref{table1} and~\ref{table3} also yield 
two predictions for the 3rd RRR pattern\footnote{A recent study of this
pattern was done in~\cite{Chk}}:
\begin{eqnarray} 
\left| V_{ub} \right| & = & \sqrt{\frac{m_u}{m_t}} \times (1+\o2)
 = 0.0036 \pm 0.0004 \; , \label{eq:Vub} \\
\left| \frac{V_{us}}{V_{cs}} \right| & = & \sqrt{\frac{m_d}{m_s}} \times (1+\o2) 
 = 0.224 \pm 0.022  \label{eq:Vus} \; .
\end{eqnarray}
While Eq.~(\ref{eq:Vus}) is scale-independent, Eq.~(\ref{eq:Vub}) is 
scale-dependent, and the number given there is at $M_Z$.
In the framework of supersymmetric grand unified theories (SUSY GUT),
pattern~3 and thus Eqs.~(\ref{eq:Vub}) and (\ref{eq:Vus}) are assumed 
to be valid at $M_X=2\times 10^{16}{\rm~GeV}$. 
Renormalization group equations (RGE) running gives $V_{ub}$
a central value $\approx 0.0033$ at $M_Z$.
The predictions are in good agreement with present data, 
approximately independent of the scale at which the pattern is valid.

  The 5th RRR pattern differs from the other four in that it allows
for two separate solutions to the Yukawa matrices with different predictions
for quark mixings. These two solutions correspond to 
$d_U>0$ (5a) and $d_U<0$ (5b).
Pattern~5 can be brought into texture VII of Table~\ref{table1} by a
common LH (12) rotation $R^{\dag}_{12}(c_U\lambda^2/e_U)$. The two predictions 
$a_U=-|c_U|^2\left(1+\frac{|e_U|^2}{d_U}\right)$ and $|b_D|^2=-a_D d_D$ can
be both written in terms of physical parameters. 
After some manipulations, we obtain
\begin{eqnarray}
\left|\frac{m_u}{m_t}\right| &=&  
\left|V_{ub}\right|^2 \Gamma_{\pm}\times (1+\o2)\; ,
\label{patt5a}\\
\left|\frac{m_d}{m_s}\right| &=&
  \left|\frac{1}{V_{ts}^*}\right|^2 
  \left|\frac{V_{ub}}{V_{cs}^*}\Gamma_{\pm}+V_{td}^*\right|^2\times (1+\o2)\; ,
\label{patt5b}\\ 
\Gamma_{\pm}^{-1}&\equiv& 
1 \pm \left|\frac{m_t}{m_c}\right|\frac{|V_{ts}^*|^2}{|V_{cs}^*|}.
\end{eqnarray}
Here the $\pm$ sign in $\Gamma_{\pm}$ corresponds to choosing the $\pm$ sign for $d_U$. Note that $\Gamma_{\pm}$ is scale-dependent.
But it can be verified that Eq.~(\ref{patt5b}) is rephasing invariant. 
Numerically, $\left|\Gamma_{\pm}^{-1}-1 \right|\simeq 0.43$ at $M_Z$.

It is interesting to note that in the limit $\Gamma_{\pm}\rightarrow 1$,
Eqs.~(\ref{patt5a},\ref{patt5b}) are reduced to the two predictions
of the 3rd RRR pattern (Eqs.~(\ref{eq:Vub},\ref{eq:Vus})):
$m_u/m_t\simeq |V_{ub}|^2$ and $m_d/m_s\simeq |V_{cd}|^2$ 
(using $V_{ub}+V_{cs}^*V_{td}^*=V_{cd}^*V_{ts}^*$).
The significant deviation of $\Gamma_{\pm}$ from unity shows that
the predictions of the 3rd and the 5th RRR patterns are, in principle,
mutually exclusive.
To obtain a numerical estimate, Eq.~(\ref{patt5a}) can be rewritten
as (including $\o2$ terms),
\be \label{eq:RRR5}
\left| \frac{V_{ub}}{V_{cb}} \right|  =
\sqrt{\frac{m_u}{m_c} \left( \frac{m_c}{m_tV^2_{cb}} \pm 1 \right)}
\times (1+\o3)
= \left\{ \begin{array}{c} 0.107 \pm 0.012 \;\;\;\; (d_U>0, {\rm~~5a})
 \\ 0.068  \pm 0.011 \;\;\;\; (d_U<0, {\rm~~5b}) 
\end{array} \right .
\ee
where the numbers are given at $M_Z$ with $V_{cb}=0.040$.
If pattern 5 is assumed to be valid at $M_X$ as in SUSY GUT, 
the central values of $V_{ub}/V_{cb}\approx  0.103$ ($d_U>0$) and $\approx 0.062$ ($d_U<0$)
for $V_{cb}\simeq 0.033$ at $M_X$.
Indeed, we see that the $V_{ub}/V_{cb}$ 
predictions of patterns 3 and 5 are significantly different.
Future measurements of $V_{ub}$ at the $B$ factories should
be able to distinguish between the two patterns. 
Note that Eq.~(\ref{patt5b}) is
sensitive to the relative phase between $V_{ub}$ and $V_{td}^*$ which is fixed
by fitting $|V_{us}|$,
and a simple estimate of $V_{td}/V_{ts}$ cannot be similarly made. 
However, the fact that $\Gamma_{\pm}$ is very different from unity, together
with Eqs.(\ref{eq:Vus},\ref{patt5b}), causes $V_{td}/V_{ts}$ to deviate 
considerably from its standard model best-fit value.

 To check the analytic predictions,
we have solved numerically the five hermitian RRR patterns using
the quark masses~\cite{RGE} and $V_{cb}=0.040$ as input parameters.
For RRR patterns 1, 2, 4, and 5, we fix the phase to get the correct
value of $V_{us}$. $V_{ub}$ and $V_{td}$ are then obtained as outputs.
For pattern 3, the phase is fixed by $V_{td}$ since $V_{us}$ as predicted 
in Eq.~(\ref{eq:Vus}) is insensitive to it. Here $V_{ub}$ and $V_{us}$ are
outputs.
The results for the CKM elements at $M_Z$ are given in Table~\ref{table4}. 
The two sets of numbers correspond to assuming the patterns to be valid at
$M_Z$ ($M_X$).
The predictions of Eqs.~(\ref{eq:Vubcb}, \ref{eq:Vtdts}, \ref{eq:Vub}, 
\ref{eq:Vus}, \ref{eq:RRR5}) are in good agreement with the exact numerical
results.
Based on the values of $V_{td}$ and $V_{ub}$ given in Table~\ref{table4}, 
it is seen 
that pattern 5a is marginal and 5b is not favored.
Note that the accuracy of the analytic predictions can be systematically 
improved by including higher order terms in $\lambda$,
assuming the quark masses are precisely known.

To summarize, RRR patterns 1, 2, and 4 give a
too small $V_{ub}/V_{cb}$, pattern 5b gives a too large $V_{td}/V_{ts}$ and 
a small $V_{ub}/V_{cb}$,
the fitting of pattern 5a with data requires a bit of stretch, and
only the 3rd RRR pattern fits the data well. This conclusion
is in disagreement
with the results of Ref.~\cite{rand}.

{\bf 4.~}{\bf Conclusions.}

Generic mass matrices contain not only the physical parameters (masses and
mixing angles), but also arbitrary right-handed rotations as well as common
left-handed rotations for both the $U$- and $D$-quark sectors. These
rotations can mask the real features of the mass matrices. 
In the minimal parameter scheme, one fixes these rotations by
imposing a sufficient number of conditions on the mass matrices. A realization
of this scheme is to transform both the $U$- and $D$-type matrices into the
upper triangular form, and demand that there be three texture 
zeros shared by the two matrices. Because of the hierarchical structures
in both the quark masses and their mixing angles, the resulting matrices
are particularly simple in the hierarchical basis. 
All of their matrix elements are simple products 
of the masses and the CKM matrix elements.
A complete classification of these matrices has been given and is listed in
Table~\ref{table1}.

Triangular matrices can be easily put into hermitian form and 
used to classify and analyze hermitian texture zeros.
Diagonal zeros in the hermitian mass matrices often manifest as simple
relations between triangular matrix elements,
which in turn imply certain relations between quark mixing angles and masses. 
By establishing the connection between triangular and hermitian matrices,
we have analyzed hermitian mass matrices with five texture zeros.
Simple, analytic predictions for quark mixings are obtained for each of
the five Ramond-Roberts-Ross  patterns.
In particular, we note that the 5th RRR pattern allows two distinctive
predictions, and that patterns 3 and 5 are mutually exclusive.
Comparison with data indicates that patterns 1, 2, and 4
are disfavored by the combined measurements of $V_{ub}/V_{cb}$,
that pattern 5 is marginally acceptable,
and that only pattern 3 fits well.

  Triangular matrices also allow one to go beyond hermitian mass matrices 
and study
texture zeros in generic mass matrices which may be associated with
certain flavor symmetries.
In fact, any  conceivable relations between quark masses and mixings
translate simply into certain relations among the triangular matrix parameters. 
A transformation of the triangular matrices to matrices in other forms 
may reveal texture zeros in the new basis.
This method can be both useful for analyzing general mass matrices,
and helpful in a model-independent, bottom-up approach to the problem of
quark masses and mixings.

{\bf Acknowledgements}

T.~K.~ and G.~W.~ are supported by the DOE, Grant no.~DE-FG02-91ER40681. S.~M.~
is supported by the Purdue Research Foundation.
G.~W.~ would like to thank the Theory Group at Fermilab where part of this work
was done.

\newpage


{
\begin{table}
\centering
\small
\caption{Classification of triangular quark mass matrices in the
hierarchical, minimal parameter basis.
The form of each pattern is invariant under phase redefinitions of
the quarks that satisfy ${\rm det}V_{\rm CKM}=1$.
The $CP$-violating measure,
$J.T.B. \simeq J m_t^4m_b^4m_c^2m_s^2$,
with $J$ being the Jarlskog parameter;
$\a_i\equiv \left|M^U_{1i}\right|$ ($i=1,2,3$),
$\b_1\equiv \left|M^U_{22}\right|$, $\b_2\equiv \left|M^U_{23}\right|$,
and $\c\equiv \left|M^U_{33}\right|$,
while $a_i\equiv \left|M^D_{1i}\right|$ ($i=1,2,3$),
$b_1\equiv \left|M^D_{22}\right|$,
$b_2\equiv \left|M^D_{23}\right|$, and $c\equiv \left|M^D_{33}\right|$.
$\alpha$, $\beta$, and $\gamma$ are angles of the unitarity triangle.
The mass matrix elements of patterns I, III, IV, VI, VIII, IX are 
accurate up to 
$1+\o4$ corrections,  those of II are corrected by
$1+\o3$, and those of V, VII, X by $1+\o2$.
}
\label{table1}
\begin{tabular}{ccccc}
Texture&$M^U$&$M^D$&$V_{\rm CKM}$&$J.T.B$\\
\hline
&&&&\\

{\rm I}&$\left(\matrix{m_u&0&0\cr 0&m_c&0\cr 0&0&m_t}\right)$&
$\left(\matrix{\frac{m_d}{V_{ud}^*}&m_sV_{us}&m_bV_{ub}\cr 0&
m_sV_{cs}&m_bV_{cb}\cr 0&0&m_bV_{tb}}\right)$&
$R^D_{23}R^D_{13}R^D_{12}$&
$ \matrix{a_2a_3b_1b_2c^2\b_1^2\c^4 \sin{\Phi_{\rm I}}\cr
(\Phi_{\rm I}={\rm arg}\left(\frac{a_2b_2}{a_3b_1}\right)\simeq\gamma)}$\\
&&&&\\


{\rm II}&$\left(\matrix{m_u&-m_c\frac{V_{ub}}{V_{cb}}&0\cr 0&m_c&0\cr 0&0&m_t}\right)$&
$\left(\matrix{\frac{m_d}{V_{ud}^*}&m_s\frac{V_{td}^*}{V_{cb}}&0\cr 0&
m_sV_{cs}&m_bV_{cb}\cr 0&0&m_bV_{tb}}\right)$&
$R^U_{12}R^D_{23}R^D_{12}$&
$\matrix{a_2b_1b_2^2c^2\a_2\b_1\c^4 \sin{\Phi_{\rm II}}\cr 
(\Phi_{\rm II}={\rm arg}\left(\frac{A_2b_1}{a_2B_1}\right)\simeq\alpha)}$\\
&&&&\\


{\rm III}&$\left(\matrix{m_u&0&-m_t\frac{V_{ub}}{V_{tb}}\cr 0&m_c&0\cr 0&0&m_t}\right)$&
$\left(\matrix{\frac{m_d}{V_{ud}^*}&m_sV_{us}&0\cr 0&m_sV_{cs}&
m_bV_{cb}\cr 0&0&m_bV_{tb}}\right)$&
$R^U_{13}R^D_{23}R^D_{12}$&
$\matrix{a_2b_1b_2c^3\a_3\b_1^2\c^3 \sin{\Phi_{\rm III}}\cr 
(\Phi_{\rm III}={\rm arg}\left(\frac{A_3b_1c}{a_2b_2C}\right)\simeq
\pi-\gamma)}$\\
&&&&\\


{\rm IV}&$\left(\matrix{m_u&0&0\cr 0&m_c&-m_t\frac{V_{cb}}{V_{tb}}\cr 0&0&m_t}\right)$
&$\left(\matrix{\frac{m_d}{V_{ud}^*}&m_sV_{us}&m_bV_{ub}\cr 0&m_sV_{cs}&
0\cr 0&0&m_bV_{tb}}\right)$&$R^U_{23}R^D_{13}R^D_{12}$&
$\matrix{a_2a_3b_1c^3\b_1^2\b_2\c^3 \sin{\Phi_{\rm IV}}\cr 
(\Phi_{\rm IV}= {\rm arg}\left(\frac{a_3b_1C}{a_2cB_2}\right)\simeq\pi-\gamma)}$\\
&&&&\\


{\rm V}&$\left(\matrix{m_u&-m_c\frac{V_{us}}{V_{cs}}&0\cr 0&m_c&0\cr 0&0&m_t}\right)$&
$\left(\matrix{\frac{m_d}{V_{ud}^*}&0&-m_b\frac{V_{td}^*}{V_{cs}}\cr 
0&m_sV_{cs}&m_bV_{cb}\cr 0&0&m_bV_{tb}}\right)$&
$R^U_{12}R^D_{23}R^D_{13}$&
$\matrix{a_3b_1^2b_2c^2\a_2\b_1\c^4 \sin{\Phi_{\rm V}}\cr 
(\Phi_{\rm V}={\rm arg}\left(\frac{a_3B_1}{b_2A_2}\right)\simeq\beta)}$\\

\end{tabular}
\end{table}
}
\newpage
\centering
\hoffset=-0.3in
\label{table1b}
\begin{tabular}{ccccc}
\hline
\hline
Texture&$M^U$&$M^D$&$V_{\rm CKM}$&$J.T.B$\\
\hline
&&&&\\


{\rm VI}&$\left(\matrix{\frac{m_u}{V_{ud}}&m_cV_{cd}^*&m_tV_{td}^*\cr 0&
m_cV_{cs}^*&m_tV_{ts}^*\cr 0&0&m_tV_{tb}^*}\right)$&
$\left(\matrix{m_d&0&0\cr 0&m_s&0\cr 0&0&m_b}\right)$&
$R^U_{12}R^U_{13}R^U_{23}$&
$ \matrix{\a_2\a_3\b_1\b_2\c^2b_1^2c^4 \sin{\Phi_{\rm VI}}\cr
(\Phi_{\rm VI}={\rm arg}\left(\frac{A_3B_1}{A_2B_2}\right)\simeq\beta)}$\\
&&&&\\


{\rm VII}&$\left(\matrix{\frac{m_u}{V_{ud}}&m_c\frac{V_{ub}}{V_{ts}^*}&0\cr 0&
m_cV_{cs}^*&m_tV_{ts}^*\cr 0&0&m_tV_{tb}^*}\right)$&
$\left(\matrix{m_d&-m_s\frac{V_{td}^*}{V_{ts}^*}&0\cr 0&m_s&0\cr 0&0&m_b}\right)$&
$R^U_{12}R^U_{23}R^D_{12}$&
$\matrix{\a_2\b_1\b_2^2\c^2a_2b_1c^4 \sin{\Phi_{\rm VII}}\cr 
(\Phi_{\rm VII}={\rm arg}\left(\frac{A_2b_1}{B_1a_2}\right)\simeq
\alpha)}$\\
&&&&\\


{\rm VIII}&$\left(\matrix{\frac{m_u}{V_{ud}}&m_cV_{cd}^*&0\cr 0&m_cV_{cs}^*&
m_tV_{ts}^*\cr 0&0&m_tV_{tb}^*}\right)$&
$\left(\matrix{m_d&0&-m_b\frac{V_{td}^*}{V_{tb}^*}\cr 0&m_s&0\cr 0&0&m_b}\right)$&
$R^U_{12}R^U_{23}R^D_{13}$&
$\matrix{\a_2\b_1\b_2\c^3a_3b_1^2c^3 \sin{\Phi_{\rm VIII}}\cr 
(\Phi_{\rm VIII}={\rm arg}\left(\frac{A_2B_2c}{B_1Ca_3}\right)\simeq
\pi-\beta)}$\\
&&&&\\


{\rm IX}&$\left(\matrix{\frac{m_u}{V_{ud}}&m_cV_{cd}^*&m_tV_{td}^*\cr 0&m_cV_{cs}^*&
0\cr 0&0&m_tV_{tb}^*}\right)$
&$\left(\matrix{m_d&0&0\cr 0&m_s&-m_b\frac{V_{ts}^*}{V_{tb}^*}\cr 0&0&m_b}\right)$&$R^U_{12}R^U_{13}R^D_{23}$&
$\matrix{\a_2\a_3\b_1\c^3b_1^2b_2c^3 \sin{\Phi_{\rm IX}}\cr 
(\Phi_{\rm IX}={\rm arg}\left(\frac{A_2Cb_2}{A_3B_1c}\right)\simeq
\pi-\beta)}$\\
&&&&\\


{\rm X}&$\left(\matrix{\frac{m_u}{V_{ud}}&0&-m_t\frac{V_{ub}}{V_{cs}^*}\cr 
0&m_cV_{cs}^*&m_tV_{ts}^*\cr 0&0&m_tV_{tb}^*}\right)$&
$\left(\matrix{m_d&-m_s\frac{V_{cd}^*}{V_{cs}^*}&0\cr 0&m_s&0\cr 0&0&m_b}\right)$&
$R^U_{13}R^U_{23}R^D_{12}$&
$\matrix{\a_3\b_1^2\b_2\c^2a_2b_1c^4 \sin{\Phi_{\rm X}}\cr 
(\Phi_{\rm X}={\rm arg}\left(\frac{B_2a_2}{A_3b_1}\right)\simeq
\gamma)}$\\
 &&&& \\
\hline
\hline
\end{tabular}
\newpage


\begin{table}
\small
\caption{The five hermitian matrices appearing in RRR patterns 
and their triangular forms.
A hierarchical matrix structure as exhibited in 
Eqs.~(\ref{eq:triUD}), (\ref{eq:HU}), and (\ref{eq:HD}) is assumed.
$A^\prime \equiv -|C|^2\left(1+\frac{|E|^2}{D}\right)$,
$D^\prime \equiv D+|E|^2$.
}
\label{table2}
\centering
\begin{tabular}{ccccc}
$M_1$ & $M_2$ & $M_3$ & $M_4$ & $M_5$\\
\hline
&&&&\\
$\left(\matrix{ 0&B&0\cr B^*&D&0\cr 0&0&1}\right)$&

$\left(\matrix{ 0&B&0\cr B^*&0&E\cr 0&E^*&1}\right)$&

$\left(\matrix{ 0&0&C\cr 0&D&0\cr C^*&0&1}\right)$&

$\left(\matrix{ 0&B&0\cr B^*&D^\prime&E\cr 0&E^*&1}\right)$&

$\left(\matrix{ 0&0&C\cr 0&D^\prime&E\cr C^*&E^*&1}\right)$\\

&&&&\\

$\left(\matrix{ -\frac{|B|^2}{D}&B&0\cr 0&D&0\cr 0&0&1}\right)$&

$\left(\matrix{ \frac{|B|^2}{|E|^2}&B&0\cr 0&-|E|^2&E\cr 0&0&1}\right)$&

$\left(\matrix{ -|C|^2&0&C\cr 0&D&0\cr 0&0&1}\right)$&

$\left(\matrix{ -\frac{|B|^2}{D} & B&0\cr 0&D&E\cr 0&0&1}\right)$&

$\left(\matrix{ A^\prime &-CE^*&C\cr 0&D&E\cr 0&0&1}\right)$\\
&&&&\\
\end{tabular}

\end{table}


\begin{table}
\small

\caption{The five RRR hermitian patterns and their corresponding triangular 
patterns of Table~I.
Also shown are the relations among the triangular matrix elements implied by
each pattern. For $M_i~(i=1,\dots,5)$, see Table~\ref{table2}. 
$a_j,\dots,e_j~ (j=U,D)$ are defined in Eq.~(\ref{eq:triUD}). }
\label{table3}
\centering
\begin{tabular}{ccccc}
RRR Pattern & $Y^U$ & $Y^D$ & Triangular Pattern & Relations\\
\hline
1& $M_1$ & $M_4$ & II & $\begin{array}{c}|b_U|^2=-a_U d_U \\ |b_D|^2=-a_D d_D
\end{array}$\\
\hline
2& $M_2$ & $M_4$ & $\begin{array}{c}{\rm II~or~VII} \\ {\rm (LH~23~rotation)}
\end{array}$ 
& $\begin{array}{c} |b_U|^2= - a_U d_U \\ |e_U|^2=-d_U \\
|b_D|^2=-a_D d_D \end{array}$\\
\hline
3& $M_3$ & $M_4$ & III & $\begin{array}{c} |c_U|^2=-a_U\\ |b_D|^2=-a_D d_D
 \end{array}$\\
\hline
4& $M_4$ & $M_1$ & VII & $\begin{array}{c} |b_U|^2=-a_U d_U \\ |b_D|^2=-a_D d_D
\end{array}$\\
\hline
5& $M_5$ & $M_1$ & $\begin{array}{c}{\rm VI, VII, or~X} \\ 
 {\rm (LH~12~rotation)} \end{array}$ 
& $\begin{array}{c} b_U=-c_U e_U^* \\ 
 |c_U|^2=-a_U/\left(1+\frac{|e_U|^2}{d_U}\right) \\
 |b_D|^2=-a_D d_D\end{array}$
\end{tabular}
\end{table}


\begin{table}
\small
\caption{
Numerical solutions to the five RRR patterns. Inputs are the central
values of quark masses 
and $V_{cb}$ and $V_{us}$ for patterns 1, 2, 4, and 5, and $V_{cb}$ and
$V_{us}$ for pattern 3. 
Cases 5a and 5b refer to $d_U>0$ and $d_U<0$, respectively. 
The two sets of numbers 
correspond to assuming
the patterns to be valid at $M_Z$ ($M_X=2\times 10^{16}{\rm~GeV}$ of SUSY GUT).
The CKM elements and $J$ shown are at $M_Z$.}
\label{table4}
\centering
\begin{tabular}{ccccccc}
Pattern & $V_{ud}$ & $V_{us}$ & $V_{ts}$ 
& $\left|\frac{V_{ub}}{V_{cb}}\right|$ & 
$\left|\frac{V_{td}}{V_{ts}}\right|$ & $J$\\
\hline
1 & \ba 0.9755\\(0.9755) \ea &  \ba 0.220\\(0.220) \ea & 
  \ba 0.0390\\(0.0391) \ea &
\ba 0.0583\\(0.0584)\ea & \ba 0.217\\(0.217)\ea &
\ba 1.9\times 10^{-5}\\(1.9\times 10^{-5}) \ea \\
\hline
2 & \ba 0.9753\\(0.9754) \ea & \ba 0.221\\(0.221) \ea &
 \ba 0.0393\\(0.0396) \ea &
\ba 0.0592\\(0.0589)\ea & \ba 0.207\\(0.209)\ea &
\ba 1.9\times 10^{-5}\\(1.9\times 10^{-5})\ea \\
\hline
3 & \ba 0.9758\\(0.9757) \ea & \ba 0.219\\(0.219) \ea &
 \ba 0.0393\\(0.0394) \ea &
\ba 0.0934\\(0.0871)\ea & \ba 0.193\\(0.193)\ea &
\ba 2.7\times 10^{-5}\\(2.6\times 10^{-5})\ea \\
\hline
4 & \ba 0.9755\\(0.9755) \ea & \ba 0.220\\(0.220) \ea &
 \ba 0.0393\\(0.0387) \ea &
\ba 0.0585\\(0.0585)\ea & \ba 0.224\\(0.223)\ea &
\ba 2.0\times 10^{-5}\\(1.9\times 10^{-5})\ea \\
\hline
5a & \ba 0.9755\\(0.9754) \ea & \ba 0.220\\(0.220) \ea &
 \ba 0.0393\\(0.0387) \ea &
\ba 0.107\\(0.103)\ea & \ba 0.238\\(0.236)\ea &
\ba 3.7\times 10^{-5}\\(3.4\times 10^{-5})\ea \\
\hline
5b & \ba 0.9754\\(0.9753) \ea & \ba 0.221\\(0.221) \ea &
 \ba 0.0391\\(0.0386) \ea &
\ba 0.0670\\(0.0607)\ea & \ba 0.245\\(0.243)\ea &
\ba 2.3\times 10^{-5}\\(2.0\times 10^{-5})\ea

\end{tabular}
\end{table}

\end{document}